%% file: RJwrapper.tex
\documentclass[a4paper]{report}
\usepackage[utf8]{inputenc}
\usepackage[T1]{fontenc}
\usepackage{RJournal}
\usepackage{amsmath,amssymb,array}
\usepackage{booktabs}

\usepackage{algorithm}
\usepackage{xcolor}
\usepackage{tabularray}
\usepackage{colortbl}
\usepackage{graphicx}
\usepackage{algpseudocode}
\usepackage{titlesec}
\renewcommand\thesection{\arabic{section}}

\def\sD{{\mathcal D}}

\def\vectorfontone{\bf}
\def\vectorfonttwo{\boldsymbol}
\def\va{{\vectorfontone a}}                      %
\def\vr{{\vectorfontone r}}                      %
\def\vy{{\vectorfontone y}}                      % Targets/Labels

\def\vone{{\vectorfontone 1}}
\def\vzero{{\vectorfontone 0}}

\def\vbeta{{\vectorfonttwo \beta}}               % Unpenalized coefficients
\def\vdelta{{\vectorfonttwo \delta}}             %
\def\matrixfontone{\bf}

\def\mA{{\matrixfontone A}}                      %
\def\mI{{\matrixfontone I}}                      % Identity Matrix
\def\mX{{\matrixfontone X}}                      % Unpenalized Design Matrix/Nullspace Matrix
\def\ds{\displaystyle}

\begin{document}

%% do not edit, for illustration only
\sectionhead{}
\volume{XX}
\volnumber{YY}
\year{20ZZ}
\month{AAAA}

%% replace RJtemplate with your article
\begin{article}
  \input{bayesianlasso}

\end{article}

\end{document}

%% file: bayesianlasso.tex
% !TeX root = RJwrapper.tex
\title{The Lasso Distribution: Properties, Sampling Methods, and Applications in Bayesian Lasso Regression}
\author{by Mohammad Javad Davoudabadi, Jonathon Tidswell, Samuel Muller,  Garth Tarr and John T. Ormerod}

\maketitle

%% I've done a bit of this, but read carefully the author guide
%% and make sure we're adhering to their specifications:
%% https://journal.r-project.org/share/author-guide.pdf
%% from that document, it's clear we need to make the focus
%% of this article more squarely on the package, but even 
% then it's not guaranteed that it will be accepted...
%% For "Add on packages"...
%% > Authors need to make a strong
%% > case (in a motivating letter accompanying a submission) for such introductions, based for
%% > example on novelty in implementation and use of R, or the introduction of new data structures
%% >representing general architectures that invite re-use

\abstract{
In this paper, we introduce a new probability distribution, the Lasso distribution. We derive several fundamental properties of the distribution, including closed-form expressions for its moments and moment-generating function. Additionally, we present an efficient and numerically stable algorithm for generating random samples from the distribution, facilitating its use in both theoretical and applied settings. We establish that the Lasso distribution belongs to the exponential family.  A direct application of the Lasso distribution arises in the context of an existing Gibbs sampler, where the full conditional distribution of each regression coefficient follows this distribution. This leads to a more computationally efficient and theoretically grounded sampling scheme. To facilitate the adoption of our methodology, we provide an R package, BayesianLasso, available on CRAN, implementing the proposed methods. Our findings offer new insights into the probabilistic structure underlying the Lasso penalty and provide practical improvements in Bayesian inference for high-dimensional regression problems.
}

%Introductory section which may include references in parentheses
%\citep{R}, or cite a reference such as \citet{R} in the text. 
\section{Introduction}
The Lasso (Least Absolute Shrinkage and Selection Operator) regression method, introduced by \citet{tibshirani1996regression}, has become a cornerstone in high-dimensional statistical modeling due to its ability to perform variable selection and regularization simultaneously. The Bayesian formulation of the Lasso has been extensively studied, often relying on a Laplace prior for the regression coefficients, which can be expressed as a scale mixture of a Gaussian distribution \citep{park2008bayesian}. However, despite its widespread use, existing formulations often face computational challenges, particularly in efficiently sampling from full conditional distributions in a Gibbs sampling framework \citep{hans2009bayesian}.

In this paper, we propose a new probability distribution, referred to as the Lasso distribution, which arises naturally in the Bayesian formulation of Lasso regression. We fully develop this distribution and derive several of its key properties, including moments, a moment-generating function, and an efficient, numerically stable method for sampling from it. 
Key to these derivations is the accurate and numerically stable evaluation of the Mill's ratio that arises in several of these functions \citep{Mills1926}. 
Further, we establish that the Lasso distribution belongs to the class of exponential family distributions, making it a theoretically attractive choice for modeling and inference. 
It is important to note that our formulation of the Lasso distribution differs from the one implemented in the \CRANpkg{LaplacesDemon} package \citep{LaplacesDemon}, which instead corresponds to the scale mixture of normals prior used in \citet{park2008bayesian}.

As an application of the Lasso distribution, we note that it arises naturally as part of a Gibbs sampling scheme, where each coefficient is sampled individually \citep[see][]{hans2009bayesian}. By leveraging the Lasso distribution as the full conditional distribution for the regression coefficients, we achieve a more computationally efficient 
%and theoretically grounded 
sampling scheme. To facilitate reproducibility and practical implementation, we provide an R package that implements our proposed methods.

The remainder of the paper is structured as follows. Section \ref{Sec:Bayes_Lasso} outlines our Bayesian hierarchical model, which motivates the Lasso distribution. In Section \ref{Sec:LassoDistribution}, we introduce the Lasso distribution. In Section \ref{sec:lasso_samples}, we describe how to sample efficiently in a numerically stable manner from the Lasso distribution. Section \ref{Sec:UsingBayesPackage} describes how to use the package \CRANpkg{BayesianLasso}. An application of the Lasso distribution via Gibbs sampling is given in Section \ref{Sec:Gibbs}. We present a performance comparison on benchmark datasets in  Section \ref{Sec:Results} and conclude with a brief discussion in Section \ref{Sec:discussion}.

%%%%%%%% The Bayesian Lasso %%%%%%%%%%%%%%%%%%%%%%%%%%%%%%%%%%%%%%%%%%%%%%%%%%%%%%%%

\section{The Bayesian Lasso}
\label{Sec:Bayes_Lasso}

We consider the standard linear regression model $\vy \sim \mathcal{N}(\mX\vbeta, \sigma^2 \mI_n)$ for the observed dataset $\sD = \{\vy, \mX \}$, where $\vy$ is an $n$-dimensional vector of centered responses, $\mX$ is an $n \times p$ matrix of standardized predictors, $\vbeta$ is a $p$-dimensional vector of regression coefficients, and $\sigma^2$ denotes the residual variance. In practice, it is often desirable to perform variable selection alongside parameter estimation, particularly when many covariates may be irrelevant. To this end, \citet{tibshirani1996regression} proposes the Lasso, which introduces an $\ell_1$ penalty to encourage sparsity in the estimated coefficients. The regularization strength is governed by a non-negative tuning parameter $\lambda$.

\begin{equation}\label{eq:aux_representation}
\beta_j| \sigma^2, \tau_j 
    \sim N( 0, \sigma^2 \tau_j),
\quad \mbox{and} \quad
\tau_j 
    \stackrel{\text{iid}}{\sim} \text{Gamma}(1,\lambda^2/2), 
\quad 1\leq j \leq p.
\end{equation}
This hierarchical representation allows for full posterior inference while promoting sparsity through the prior structure. Here, we adopt independent conjugate priors for $\sigma^2$ and $\lambda^2$: $
\sigma^2 \sim \text{IG}(A, B)$
and $\lambda^2 \sim \mbox{Gamma}(u,v)$
where $A>0$, $B>0$, $u>0$, and $v>0$ are fixed prior hyperparameters.

We propose a modification to the hierarchical prior model in \eqref{eq:aux_representation}, which enables a kernel representation of the full conditional distribution of each $\beta_j$, as given in \eqref{LassoProp}, through the following alternative specification:
\begin{equation}\label{equ3}
\beta_j | \sigma^2, a_j 
    \sim N\left( 0, \frac{\sigma^2}{a_j\lambda^2} \right),
\quad \text{and} \quad
a_j \stackrel{\text{iid}}{\sim} \text{IG}(1, 1/2), \quad 1 \leq j \leq p.
\end{equation}

\citet{hans2009bayesian} takes a different approach. Instead of using the auxiliary variable representations in \eqref{eq:aux_representation} and \eqref{equ3}, the standard Gibbs sampler of \citet{hans2009bayesian} samples from the full conditional distributions of the parameters individually, employing a weighted combination of two truncated normal distributions. It can be shown that the full conditional distribution for each $\beta_j$ is proportional to 
\begin{align}\label{LassoProp}
p(\beta_j|\sD,\vbeta_{-j},\sigma^2,\lambda^2) \propto 
\exp\left(-\tfrac{1}{2}a\beta_j^2 + b\beta_j - c |\beta_j|\right)
\end{align}
where $a$, $b$, and $c$ are constants. This kernel corresponds to the weighted combination of two truncated normal distributions utilized by \citet{hans2009bayesian}. Further details can be found in the Supplementary Material.

Notably, this kernel does not correspond to a well-known probability distribution that can be directly sampled within a standard Gibbs sampling framework. To address this challenge, we develop a new distribution, which we term the {\em Lasso distribution}, as it arises naturally in the context of the Lasso regression model. In the next section, we formally define the Lasso distribution and derive several of its key properties, including its probability density function (PDF), cumulative distribution function (CDF), moment generating function (MGF), and moments.

%%%%%%%% The Lasso Distribution %%%%%%%%%%%%%%%%%%%%%%%%%%%%%%%%%%%%%%%%%%%%%%%%%%%%%%%%%%%%%%%%%%%%%%%%%%%%%%%%%%%%

\section{The Lasso distribution}\label{Sec:LassoDistribution}

We derive several properties of the Lasso distribution, including the probability density function (PDF), cumulative density function (CDF), moment generating function (MGF), and moments. For detailed derivations, please refer to Section \ref{Suppl:deriv_LassoDist_properties} of the Supplementary Materials.

% % In this study, we introduce a novel distribution specifically tailored for the Gibbs sampler as the full conditional distribution of the regression coefficients within the framework of the Lasso regression. Our goal is to leverage this distribution to facilitate shrinkage and improve the performance of the Gibbs sampler in estimating the regression coefficients.

% The introduced distribution is designed to incorporate shrinkage properties, allowing for more effective regularization and variable selection. By utilizing this distribution within the Gibbs sampler, we can draw samples from it to update the regression coefficients iteratively. This approach ensures that the resulting estimates exhibit desirable shrinkage properties, promoting sparsity and reducing the impact of irrelevant variables.

A random variable $X$ that maps from a probability space to the real line $\mathbb{R}$ has a Lasso distribution with parameters $a$, $b$, and $c$, denoted $X \sim \text{Lasso}(a,b,c)$, if its PDF is 
\begin{align*}
   p(X,a,b,c) = Z^{-1}\exp\left(-\tfrac{1}{2}aX^2+bX-c|X|\right),
\end{align*}
where $a \geq 0$, $b \in \mathbb{R}$, $c \geq 0$ (with $a$ and $c$ not both 0), and normalizing constant  $Z$   given by
$$Z(a,b,c) = \int_{-\infty}^\infty \exp\left( -\tfrac{1}{2}aX^2 + bX - c|X| \right) dx = \sigma \left[ \frac{\Phi(\mu_1/\sigma)}{\phi(\mu_1/\sigma)}
+ \frac{\Phi(-\mu_2/\sigma)}{\phi(-\mu_2/\sigma)}  \right],$$
with $\mu_1 = (b-c)/a$, $\mu_2 = (b + c)/a$, and $\sigma^{2} = 1/a$. Here, $\Phi(\cdot)$ and $\phi(\cdot)$ denote the CDF and PDF of the standard normal distribution, respectively. Evaluating $Z$ can be numerically challenging due to potential overflow, underflow, or divide-by-zero problems. To address this, our R package \CRANpkg{BayesianLasso} implements a numerically stable and efficient method for computing $Z$, automatically managing these numerical difficulties \citep{BayesianLassoPkg}.

%$Z$ is a function of the Mill's ratio, $m(x) = (1 - \Phi(x))/\phi(x)$. 

%We will address these problems in Section \ref{sec:lasso_samples} of the Supplementary materials.

\begin{figure}[ht]
    \centering
    \includegraphics[width=15cm, height=12.2cm]{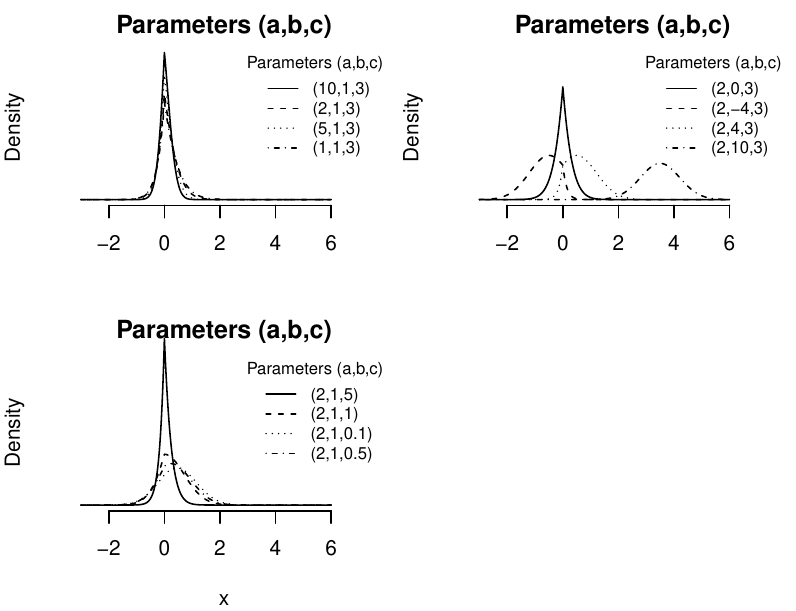}
    \vspace*{-0.5cm}
    \caption{The Lasso density function is depicted for different parameter values. In the top left panel, parameter $a$ is varied while the other parameters are fixed. In the top right and bottom left panels, parameters $b$ and $c$ are varied, respectively, while the other parameters are fixed.}
    \label{fig:density_Lassos}
\end{figure}

The Lasso distribution can be written as a mixture of two truncated normal distributions.
As we shall see, for different extreme values of $a$, $b$ and $c$, the Lasso distribution can behave like a normal distribution, a Laplace distribution, an asymmetric Laplace distribution, a positively truncated normal distribution, or a negatively truncated normal distribution. 
By characterizing the limiting behavior under extreme parameter values, the R package \pkg{BayesianLasso} leverages appropriate approximations to known distributions ensuring robust and efficient computation even in numerically challenging settings.
Figure \ref{fig:density_Lassos} shows the density of the univariate Lasso distribution for different parameter values. 

% Here, the log-partition function is
% \begin{align*}
%     A(\theta ) = \log(Z) &= \log(\int h(x)\exp(\nu(\theta) T(x))dx)\\&
%     =\log(\int \exp(-\frac{1}{2}ax^2+bx-c|x|)dx).
% \end{align*}

The MGF of the Lasso distribution is $M(t) = Z(a,b+t,c)/Z(a,b,c)$ and the calculation of the moments of the Lasso distribution via the MGF is algebraically cumbersome. See Section \ref{Suppl:deriv_LassoDist_properties} of the Supplementary Materials for the derivation of the MGF. However, by expressing the Lasso distribution as a mixture of two truncated normal distributions, the moments can be calculated as $
    \mathbb{E}(X^r) = (1 - w) \cdot \mathbb{E}(A^r) + w  \cdot \mathbb{E}(B^r),
$ where the mixing proportion is given by
$$
w =  \left[ 1
+ \frac{\Phi(\mu_1/\sigma)}{\phi(\mu_1/\sigma)}\frac{\phi(-\mu_2/\sigma)}{\Phi(-\mu_2/\sigma)}
\right]^{-1}, 
$$
where $A\sim TN_+(\mu_1,\sigma^{2})$ and $B\sim TN_-(\mu_2,\sigma^{2})$ are the positively truncated and negatively truncated normal distributions, respectively. 

The CDF of the Lasso distribution is given by
$$
\displaystyle 
P(x) = \mathcal{P}(X < x) 
     = \left\{ \begin{array}{ll}
\displaystyle 
w\frac{\Phi\left( 
\frac{x - \mu_2}{\sigma}\right)}{\Phi(-\mu_2/\sigma)} 
    & \mbox{if $x\le 0$ and}
\\ [2ex]
\displaystyle  1 - (1 - w)\frac{\Phi(\frac{\mu_1 - x}{\sigma})}{\Phi(\mu_1/\sigma)}
    & \mbox{if $x> 0$}.
\end{array} 
\right.
$$
 
The inverse CDF of the Lasso distribution is given by
$$
P^{-1}(u) = \left\{
\begin{array}{ll}
\mu_2 + \sigma\Phi^{-1}[\Phi(-\mu_2/\sigma) u/w] 
    & \mbox{if $u\le w$ and} 
\\ [2ex]
\mu_1 - \sigma\Phi^{-1}[\Phi(\mu_1/\sigma) (1 - u)/(1-w) ]
    & \mbox{if $u> w$}.
\end{array} 
\right. 
$$

Finally, the mode of the Lasso distribution is equal to $\mbox{max}(|b| - c,0)\cdot\mbox{sign}(b)/a$, where $\mbox{sign}(\cdot)$ is the sign function. 

The R package \pkg{BayesianLasso} enables efficient and numerically stable sampling from the Lasso distribution by evaluating the inverse CDF using uniform random draws. To ensure robustness, particularly in the tails, the implementation incorporates safeguards against numerical issues such as overflow and underflow. A key component of this approach is the numerically stable evaluation of the Mill’s ratio, defined as $m(x) = \Phi(-x)/\phi(-x)$, which is used in computing the normalizing constant. Specifically, the package computes 
$$
v_{1}=\frac{c-\left|b\right|}{\sqrt{a}} 
\qquad 
\mbox{and} \qquad  
v_{2}=\frac{c+\left|b\right|}{\sqrt{a}},
$$ 
and evaluates the normalizing constant as
$Z = \left[m\left(v_{1}\right)+m\left(v_{2}\right)\right]/\sqrt{a}$. This combination of numerical stability and computational efficiency underpins the practical value of the package and is a key reason for its robustness in high-dimensional Bayesian inference.

%%%% Sampling from the Lasso distribution  %%%%%%%%%%%%%%%%%%%%%%%%%%%%%%%%%%%%%%%%%%%%%%%%%%%%%%%%%%%%%%%%%%%%%%%%%%%%%%%%%%%%

\subsection{Sampling from the Lasso distribution}
\label{sec:lasso_samples}

Key quantities required for various expressions in the previous section include the computation of $Z$ and $w$. Additionally, special care must be taken when calculating the inverse CDF to prevent both overflow and underflow. To address these concerns, we define $
v_{1}=\left[c-\left|b\right|\right]/\sqrt{a}$ and $ 
v_{2}=\left[c+\left|b\right|\right]/\sqrt{a}$. The normalizing constant 
$Z$ can then be expressed as
$Z = \left[m\left(v_{1}\right)+m\left(v_{2}\right)\right]/\sqrt{a}$.
For negative inputs, i.e., $x<0$, the Mill's ratio is 
$m\left(x\right)=\phi\left(x\right)^{-1}-m\left(-x\right)$. Since $v_{2}\ge 0$ and $v_{1}\ge 0$ when $\left|b\right|\le c$, we can partition the definition of $Z$ into two cases
\begin{align*}
Z & =\begin{cases}
\frac{1}{\sqrt{a}}\left[m\left(v_{1}\right)+m\left(v_{2}\right)\right] & \left|b\right|\le c \quad \mbox{and}\\
\frac{1}{\sqrt{a}}\left[\left(\phi\left(v_{1}\right)^{-1}-m\left(-v_{1}\right)\right)+m\left(v_{2}\right)\right] & \left|b\right|>c;
\end{cases}
\end{align*}
where both alternatives have the requirement for two Mill's ratio
calculations for positive parameters; this is a potential use case
for a SIMD, which is short for single instruction/multiple data, vectorized implementation of the positive side of the Mill's ratio. 

The second component of $Z$ (for $\left|b\right|>c$) has an obvious
potential overflow when $\left|v_{1}\right|$ is large.  In such cases, it may be necessary to transition to logarithmic space to represent $\phi\left(v_{1}\right)$
to avoid underflow/overflow.
However, unnecessary logarithmic computations should be avoided, as they are computationally expensive and can reduce numerical precision when not required. \citet{marsaglia2004evaluating} notes that the precision of evaluating the standard normal density function $\phi\left(x\right)$ and its tail probability deteriorates as 
$x$ increases, due to the limitations of computing $\exp(-x^2/2)$ in double-precision arithmetic. While the paper primarily examines values up to $x \approx 16$, it is also known that $\phi\left(x\right)$ effectively loses all precision for $x\geq 37$ , where the exponential term underflows to zero and the reciprocal overflows. Now, consider the case where $v_{1}<-37$, making overflow a concern. 
By definition, we have $37<-v_{1}<v_{2}$, which implies $m\left(37\right)>m\left(-v_{1}\right)>m\left(v_{2}\right)>0$. Since $m\left(37\right)\approx0.027$, it follows that  $-0.027<m\left(v_{2}\right)-m\left(-v_{1}\right)<0$. Within the limits of double precision, this difference becomes negligible due to rounding errors for $v_{1}\approx-9$, and certainly for $v_{1}<-37$, where $\phi\left(-37\right)<6\times10^{-300}$. Consequently, $Z$ can be approximated as
\begin{align*}
Z & \approx\frac{1}{\sqrt{a}}\left[\phi\left(v_{1}\right)^{-1}\right].
\end{align*}
 Thus, we can safely use the approximation $m\left(v_{1}\right)\approx\phi\left(v_{1}\right)^{-1}$ in such cases.

For a fast and accurate approximation of Mill’s ratio for positive arguments, we employ a degree $\left(8,9\right)$ rational polynomial optimized using the Remez algorithm \citep{reemtsen1990modifications} over the interval $\left[0,600\right]$, achieving an accuracy of up to 12 significant figures. The leading coefficients of both the numerator and denominator are positive, ensuring that the approximation asymptotically converges to a small fraction. This stability allows the relative error to remain consistent, preserving 11  significant figures up to approximately $x\approx2000$. Beyond this point, the precision gradually improves, as Mill’s ratio satisfies $m(x) \sim 1/x$ for large $x$.

Let 
$$
\begin{array}{c}
p_0 = 46697.7602201933, \qquad  p_1 = 69339.6909002865, \qquad p_2 =50590.6980372328, \\
p_3 = 23184.62760379742, \qquad p_4 = 7236.31450136984, \qquad p_5 = 1572.136841909630, \\ 
p_6 = 232.9967987466022, \qquad p_7 = 21.74833514806325, \qquad \mbox{and} \quad p_8 = 1.000000000000095.
\end{array}
$$
and let
$$
\begin{array}{c}
q_0 = 37259.42190376593, \qquad q_1 = 85053.78630172011, \qquad q_2 = 89598.92885811838, \\
q_3 = 57370.93777717682, \qquad q_4 = 24713.27114352290, \qquad q_5 = 7467.311205544661, \\ 
q_6 = 1593.885178714749, \qquad q_7 = 233.9967987305447, \qquad q_8 = 21.74833514813385, \\
\end{array}
$$
and $q_{9} = 1$. Let
$$
\displaystyle \widetilde{m}  = \frac{
    p_0 + x(p_1 + x(p_2 + x( p_3 + x(p_4 + x(p_5 + x(p_6 + x(p_7 + xp_8)))))))
}{
    q_0 + x(q_1 + x(q_2 + x(q_3 + x(q_4 + x(q_5 + x(q_6 + x(q_7 + x(q_8 + x))))))))
}.
$$
Our approximation of $m(x)$ for $x>0$ is then
$$
m(x) = \left\{\begin{array}{ll}
\widetilde{m}    & \mbox{$x < 1.75 \times 10^{34}$} \\
1/x  & \mbox{otherwise.}
\end{array}\right.
$$

Next, we define 
$$
w_{1}=\frac{1}{1+m\left(v_{1}\right)m\left(v_{2}\right)^{-1}}
\qquad 
\mbox{and}
\qquad 
w_{2}=\frac{1}{1+m\left(v_{1}\right)^{-1}m\left(v_{2}\right)}.
$$
As $v_{1}\rightarrow-\infty$, it follows that $m\left(v_{1}\right)\rightarrow\infty$, leading to $w_{1}\rightarrow0$ and $w_{2}\rightarrow1$.
However, due to the dynamic range limitations of floating-point representation, $w_{1}$
is representable, whereas  $w_{2}$ rounds to 1 for $v_{1}\lessapprox-5.5$,
resulting in a loss of precision. Since floating-point arithmetic offers much better precision near zero than near one, using $w_{2}$ when $v_{1}\ll0$ can introduce numerical errors. To mitigate this, we define
$$
w=\begin{cases}
w_{1} & \mbox{if } b\le0 \mbox{ and} \\
w_{2} & \mbox{if } b>0,
\end{cases}
\qquad \mbox{and}
\qquad 
1-w =\begin{cases}
w_{2} & \mbox{if } b\le0 \mbox{ and} \\
w_{1} & \mbox{if } b>0.
\end{cases}
$$
Thus, a simple yet numerically stable rule is to work with $w$ when $b\le0$ and with $1-w$
when $b>0$.

The computation of $P^{-1}(u)$ is carried out in four distinct cases, determined by whether $u \leq w$ and whether $b > 0$. First, the arguments of $\Phi^{-1}(\cdot)$ are carefully computed to prevent overflow. In extreme cases, underflow may still occur. When underflow is detected, the arguments of $\Phi^{-1}(\cdot)$ are instead computed on the logarithmic scale, and the asymptotic tail formula of \citet{Martin} is used to evaluate $\Phi^{-1}(\cdot)$ accurately.

%%%%%% Code example of the Lasso distribution %%%%%%%%%%%%%%%%%%%%%%%%%%%%%%%%%%%%%%%%%%%%%%%%%%%%%%%%%%%%%%%%%%%%%%%%%%

\section{Using the BayesianLasso package}\label{Sec:UsingBayesPackage}

To illustrate the functionality of the \pkg{BayesianLasso} package, we provide examples of its key functions, including density evaluation, cumulative distribution, random sampling, and quantile calculations.

\subsection{Probability density function (PDF)}

The function \code{dlasso()} computes the probability density function of the Lasso distribution for given parameters. The following example plots the density of a $\text{Lasso(2,1,3)}$ distribution, shown as a solid black line in Figure \ref{fig:density_comparison}.

\begin{example}
library(BayesianLasso)

# Define parameters
a <- 2
b <- 1
c <- 3
x <- seq(-3, 3, length = 1000)

# Plot the density of Lasso(2,1,3)
plot(x, dlasso(x, a, b, c, logarithm = FALSE), type = 'l', 
     xlab = "x", ylab = "Density", 
     ylim = c(0, 2.25), yaxt = "n", bty = "n")
\end{example}

\subsection{Cumulative distribution function (CDF)}
The function \code{plasso()} calculates the cumulative distribution function of the Lasso distribution. Below, we compute the CDF at $x = -1$ for a $\text{Lasso}(2,1,3)$ distribution:

\begin{example}
CDF_value <- plasso(-1, a, b, c)
print(CDF_value)
[1] 0.00176594
\end{example}

\subsection{Random sampling from the Lasso distribution}
\begin{figure}[ht]
    \centering
    \includegraphics[width=10cm, height=8cm]{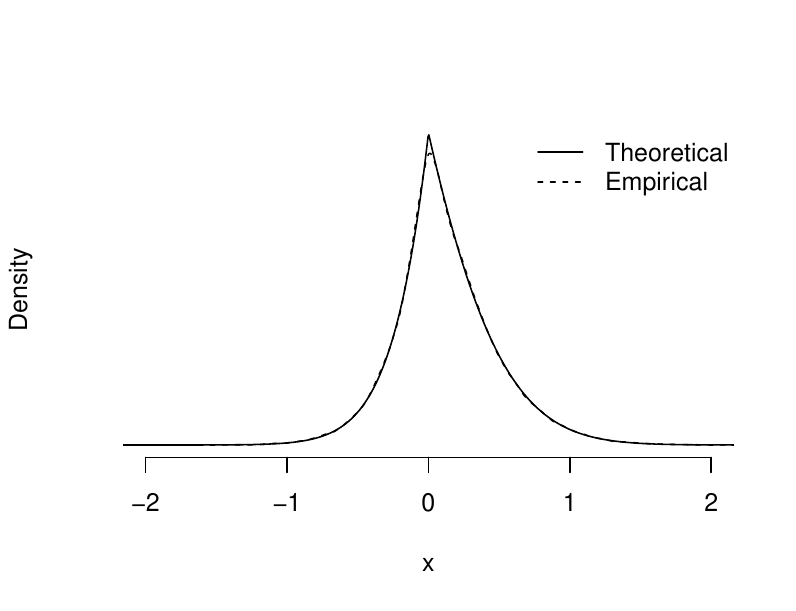}
    \vspace*{-0.5cm}
    \caption{Theoretical and empirical densities for $\text{Lasso}(2,1,3)$. The theoretical density plot of $\text{Lasso}(2,1,3)$ using \code{dlasso(x, a, b, c, logarithm = FALSE)} is shown as a solid black line, and the empirical density is shown as a dashed line.}
    \label{fig:density_comparison}
\end{figure}

To generate random samples from the Lasso distribution, we use \code{rlasso()}. The following example generates 100,000 samples and compares the empirical density with the theoretical density, as shown in Figure \ref{fig:density_comparison}.

\begin{example}
# Generate random samples
samples <- rlasso(100000, a, b, c)

# Compare empirical and theoretical density
plot(x, dlasso(x, a, b, c, logarithm = F), type = 'l', xlab = "x", ylab = "Density",
     xlim = c(-2, 2), yaxt = "n", bty = "n")
lines(density(samples), col = 'red')
legend("topright", legend = c("Theoretical", "Empirical"), 
       col = c("black", "red"), lty = 1, bty = "n")
\end{example}

\subsection{Quantile function}
The \code{qlasso()} function computes quantiles of the Lasso distribution. Below, we compute the quantiles for probability values \textbf{$p = \{0.1, 0.3, 0.6\}$}:

\begin{example}
p_values <- c(0.1, 0.3, 0.6)
quantiles <- qlasso(p_values, a, b, c)
print(quantiles)
[1] -0.28183916 -0.04935763  0.16137104
\end{example}

In the next section, we incorporate the Lasso distribution into a Gibbs sampling algorithm as the full conditional distribution for the coefficients of the Lasso regression model.

%%%%% Application of the Lasso Distribution in Gibbs Sampling  %%%%%%%%%%%%%%%%%%%%%%%%%%%%%%%%%%%%%%%%%%%%%%%%%%%%%%%%%%%%%%%%%%%%%%%%%%%%%%%%%%%%

\section{Application of the Lasso distribution in Gibbs sampling}
\label{Sec:Gibbs}

We modify the Gibbs sampler of \citet{hans2009bayesian} (henceforth Hans) by using the Lasso distribution as the full conditional distribution of the regression coefficients. We also consider a slightly modified version of the Gibbs sampler of \citet{park2008bayesian} (henceforth PC), as the alternative Gibbs sampler, by changing the representation of the auxiliary variable from \eqref{eq:aux_representation} to \eqref{equ3}.

Furthermore, these Gibbs samplers are based on the prior $\lambda^2\sim\mbox{Gamma}(u,v)$ which differs from
\citet{hans2009bayesian} which places this prior on $\lambda$ 
(which is conjugate) rather
than $\lambda^2$ (which is not). 
Therefore, we employ a slice sampler to draw from the full conditional distribution of $\lambda^2$ \citep{neal2003slice}. All methods are implemented in the same programming language and executed on the same computer, ensuring that the primary distinction lies in the choice of Gibbs samplers used to fit the model.

\subsection{Hans Gibbs sampler}

As mentioned earlier, the standard Gibbs sampler proposed by \citet{hans2009bayesian} relies on the fact that the full conditional distribution for each $\beta_j$ is given by \eqref{LassoProp}, and \citet{hans2009bayesian} 
notes that $p(\beta_j|\sD,\vbeta_{-j},\sigma^2,\lambda^2)$ can be represented as a mixture of two truncated normal distributions. This is not a well-known distribution to sample. However, we show care needs to be taken to avoid numerical problems using this representation.

To facilitate sampling from the kernel in \eqref{LassoProp}, we utilize the Lasso distribution ($\mbox{Lasso}(a,b,c)$), which we introduced in Section \ref{Sec:LassoDistribution}. We also employ the efficient and numerically stable sampling method developed in Section \ref{sec:lasso_samples} to draw samples from this distribution. Notably, while the Hans sampler is specifically designed for the Lasso distribution, it is difficult to extend to other response types or alternative penalty structures.

Algorithm~\ref{B_Gibbs} presents our modified version of Hans’ Gibbs sampling algorithm, where we model $\lambda^2 \sim \mbox{Gamma}(u,v)$ instead of $\lambda \sim \mbox{Gamma}(u,v)$. Unlike the original method in \citet{hans2009bayesian}, which employs conjugate sampling for $\lambda$ and rejection sampling for $\sigma^2$, our approach collapses the auxiliary variable $a_j$ to improve mixing and utilizes slice sampling for both $\sigma^2 \mid \sD, \vbeta, \lambda^2$ and $\lambda^2 \mid \sD, \vbeta, \sigma^2$ \citep{neal2003slice}. Notably, the same slice sampler is applied to both parameters, as their full conditional distributions are inverse transformations of each other.

\begin{algorithm}[!ht]
\caption{Modified Hans Gibbs sampler}\label{B_Gibbs}
\begin{algorithmic}[1]

\State \textbf{Inputs}: $\vy$, $\mX$, $\widetilde{a}>0$, $\widetilde{b}>0$, $u>0$, $v>0$, $\vbeta^{(0)} = \vzero_p$, $\sigma^{2(0)}>0$, $\lambda^{(0)}>0$; 

\For {$i = 1,\ldots,\textit{N}$}

\State $\vbeta^{(i)} \leftarrow \vbeta^{(i-1)}$;

\If{$n>p$} 
    \State $\vdelta \leftarrow  (\mX^T\mX) \vbeta^{(i)}$;
    \For {$j = 1,\ldots,\textit{p}$} 
        \State $\vdelta \leftarrow  \vdelta - \mX_j \beta_j^{(i)}$;
        \State $\ds \beta^{(i)}_j \sim 
            \text{Lasso}\left( 
                \frac{\| \mX_j\|_2^2}{\sigma^{2(i-1)}}, 
                \frac{(\mX^T\vy)_j - \delta_j}{\sigma^{2(i-1)}}, 
                \frac{\lambda^{(i-1)}}{\sigma^{(i-1)}} 
            \right);$
        \State $\vdelta \leftarrow  \vr + \mX_j \beta_j^{(i)}$;
    \EndFor
    \State $\mbox{RSS} \leftarrow  \|\vy\|_2^2 - 2\vy^T\mX\vbeta_j^{(i)} + (\vbeta_j^{(i)})^T\mX^T\mX\vbeta_j^{(i)}$;
\Else 
    \State $\vy \leftarrow \mX\vbeta^{(i)}$;

\For {$j = 1,\ldots,\textit{p}$} 
\State $\vr_j \leftarrow \vy - (\widehat{\vy} - \mX_{j}\beta^{(i-1)}_j);$

\State $\ds \beta^{(i)}_j \sim 
    \text{Lasso}\left( 
        \frac{\| \mX_j\|_2^2}{\sigma^{2(i-1)}}, 
        \frac{\mX_j^T\vr_j}{\sigma^{2(i-1)}}, 
        \frac{\lambda^{(i-1)}}{\sigma^{(i-1)}} 
    \right);$

\State $\widehat{\bf y} \leftarrow \widehat{\bf y} + {\bf X}_{j}(\beta^{(i)}_j - \beta^{(i-1)}_j) $;
\EndFor

    \State $\mbox{RSS} \leftarrow  \|\vy - \widehat{\vy}\|_2^2$;

\EndIf

\State Use slice sampling to sample $\sigma^{2(i)}$ from the density
$$
\ds p(\sigma^2|\sD,\vbeta^{(i)},\lambda^{2(i-1)})      
    \propto \exp\left[
        (\widetilde{a} +\tfrac{n+p}{2} +1)\log(\sigma^{2}) 
        - \left(\widetilde{b} + \tfrac{1}{2}\mbox{RSS}\right)\sigma^{-2} - \tfrac{\lambda^{(i-1)}}{\sigma}\| \vbeta^{(i)} \|_1 \right];
$$

\State Use slice sampling to sample $\lambda^{2(i)}$ from the density
$$
\ds p(\lambda^{2}|\sD,{\boldsymbol\beta}^{(i)},\sigma^{2(i)}) 
    \propto \exp\left[
        \left( u + \tfrac{p}{2} - 1\right) \log(\lambda^{2}) 
        - v\lambda^{2(i-1)} 
        - \tfrac{\lambda}{\sigma^{(i)}} \| \vbeta^{(i)} \|_1 \right].
$$

\EndFor
\end{algorithmic}
\end{algorithm}

In Algorithm \ref{B_Gibbs}, the quantity ${\bf r}_j$ is analogous to the partial residuals used in coordinate descent methods for optimizing the Lasso objective function \cite[see, e.g.,][Section 5.4.2]{HasteEtAl2015}. However, rather than computing the mode of $\beta_j|{\mathcal D},{\boldsymbol\beta}_{-j},\sigma^2$ as done in coordinate descent, we draw samples from its distribution.

\citet{hans2009bayesian} does not discuss the computational costs of the Gibbs samplers presented there. In contrast, our Algorithm \ref{B_Gibbs} avoids matrix inversion, and as shown in Algorithm \ref{B_Gibbs}, sampling each variable requires $O(\min(n,p))$  time. Moreover, Algorithm \ref{B_Gibbs} accommodates both the $n>p$ and $p>n$  settings, running in $O(N p \min(n,p))$ time.
This complexity suggests that Algorithm \ref{B_Gibbs} may be computationally more efficient than Algorithm \ref{Or_Gibbs} for datasets where $p>n$ or where $n>p$ but $n$ is of smaller order than $p^2$.

Note that our \pkg{BayesianLasso} package implements the modified Hans sampler via the function
\texttt{Modified\_Hans\_Gibbs()}. In this function, $\mX$ and $\vy$ represent the covariate matrix and the response vector, respectively. The arguments \texttt{a1}, \texttt{b1}, \texttt{u1}, and \texttt{v1} correspond to the hyperparameters of the priors for $\sigma^2$ and $\lambda^2$. The total number of MCMC iterations is specified by \texttt{nsamples}, and the initial values for $\vbeta$, $\sigma^2$ and $\lambda^2$ are set via {\tt beta\_init}, \texttt{sigma2\_init} and \texttt{lambda\_init}, respectively. The argument \texttt{verbose} controls whether the sampling progress is printed during execution.

\subsection{PC Gibbs sampler}

Our modified PC Gibbs sampler is presented in Algorithm \ref{Or_Gibbs}, which consists of a four-block Gibbs sampling scheme for estimating the model parameters, auxiliary variables, and the tuning parameter $\lambda^2$.  Specifically, it involves sampling from the full conditional distributions:
$[\vbeta|\sD,\sigma^2,\lambda^2,\va]$, 
$[\sigma^2|\sD,\vbeta,\lambda^2,\va]$, 
$[\lambda^2|\sD,\vbeta,\sigma^2,\va]$, and 
$[\va|\sD,\vbeta,\sigma^2,\lambda^2]$.
The time complexity of Algorithm \ref{Or_Gibbs} is $O(np^2 + Np^3)$ where $N$ is the number of MCMC samples, assuming that quantities such as $\mX^T\mX$, $\mX^T\vy$, and $\|\vy\|_2^2$ are precomputed outside the main loop. A key advantage of this algorithm is that the inner loop’s time complexity is independent of $n$, making it particularly beneficial when $n$ is large. However, the algorithm scales cubically with $p$, as it requires computing a matrix square root—typically via Cholesky factorization or eigen-decomposition—to sample $\vbeta^{(i)}$ \citep{golub13}.

\begin{algorithm}[!ht]
\caption{Modified PC Gibbs sampler}\label{Or_Gibbs}
\begin{algorithmic}[1]

\vspace{1mm}
\State \textbf{Inputs}: $\vy$, $\mX$, $\widetilde{a}>0$, $\widetilde{b}>0$, $u>0$, $v>0$, $\lambda^2>0$, and $\sigma^{2(0)}>0$. 

\vspace{1mm}
\State $\va^{(0)} = \vone_n$;
\For {$i = 1,\ldots,\textit{N}$}
\State $\mA \leftarrow \text{diag}(\va^{(i-1)})$;

\vspace{1mm}
\State $\vbeta^{(i)} \sim N_p\left[
    (\mX^T\mX + \lambda^{2(i-1)}\mA)^{-1}\mX^T\vy,
    \sigma^{2(i-1)}(\mX^T\mX + \lambda^{2(i-1)}\mA)^{-1}
\right]$;  

\vspace{1mm}
\State $\sigma^{2(i)} \sim \text{IG}\left(
    \widetilde{a} + \tfrac{n+p}{2},
    \widetilde{b} + \tfrac{1}{2}\| \vy \|_2^2 
      - \vy^T\mX \vbeta^{(i)}
      + \tfrac{1}{2}(\vbeta^{(i)})^T(\mX^T\mX + \lambda^{2(i-1)}\mA) \vbeta^{(i)} \right)$;

\vspace{1mm}
\State $\ds \lambda^{2(i)} \sim \text{Gamma}\left(
    u+\frac{p}{2},
    v + \frac{(\vbeta^{(i)})^T\mA\vbeta^{(i)}}{2\sigma^{2(i)}} 
\right)$;

\vspace{1mm}
\For {$j = 1,\ldots,\textit{p}$}  
\State $\ds a^{(i)}_j \sim \text{Inverse-Gaussian}\left( \frac{\sigma^{(i)}}{\lambda|\beta_j^{(i-1)}|},1 \right)$.
\EndFor

\EndFor
\end{algorithmic}
\end{algorithm}
Note that our \pkg{BayesianLasso} package implements the modified PC sampler via the function
\texttt{Modified\_PC\_Gibbs()}. In this function, $\mX$ and $\vy$ represent the covariate matrix and the response vector, respectively. The arguments \texttt{a1}, \texttt{b1}, \texttt{u1}, and \texttt{v1} correspond to the hyperparameters of the priors for $\sigma^2$ and $\lambda^2$. The total number of MCMC iterations is specified by \texttt{nsamples}, and the initial values for $\sigma^2$ and $\lambda^2$ are set via \texttt{sigma2\_init} and \texttt{lambda\_init}. The argument \texttt{verbose} controls whether the sampling progress is printed during execution.

%%%%%%%%%  Results %%%%%%%%%%%%%%%%%%%%%%%%%%%%%%%%%%%%%%%%%%%%%%%%%%%%%%%%%%%%%%%%%%%%%%%%%%%%%%%%%%%%

\section{Performance comparison on benchmark datasets}
\label{Sec:Results}

In this section, we compare the performance of our modified Hans sampler with our modified PC Gibbs sampler, as well as the R packages \CRANpkg{monomvn}, \CRANpkg{bayeslm}, \CRANpkg{rstan}, and \CRANpkg{bayesreg} \citep{monomvn,bayeslm,Stan_Development_Team2020-sz,makalic2016high}. The comparison is based on several benchmark datasets that represent a diverse range of scenarios where $n>p$. Specifically, we consider the Diabetes dataset with all pairwise interactions of the original variables (referred to as Diabetes${}^2$) from the \CRANpkg{lars} package \citep{efron2004least, lars}, with $n = 442$ and $p = 55$;  the Kakadu dataset with all pairwise interactions (Kakadu${}^2$) from the \CRANpkg{Ecdat} package \citep{Ecdat}, with $n = 1827$ and $p = 252$; and the Crime dataset from the UCI Machine Learning Repository with $n = 2215$ and $p = 98$ \citep{communities_and_crime_183}.

The impact of autocorrelation within the chains on estimation uncertainty can be quantified using the effective sample size (ESS). We compute ESS using the \code{ess\_bulk()} function from the \CRANpkg{posterior} package in R \citep{BurknerEtAl2023,VehtariEtAl2021}.
To evaluate the efficiency of our proposed MCMC approach relative to the modified PC Gibbs sampler and the aforementioned R packages, we use the following metric
$$
\mbox{Efficiency} = \frac{\mbox{ESS}}{\mbox{time}}; 
$$
where $\mbox{time}$ represents execution time in seconds. Additionally, we assess whether the MCMC samples have reached a stationary distribution and exhibit adequate mixing using Gelman and Rubin’s convergence diagnostic, $\widehat{R}$ \citep{gelman1992inference}. We assess whether the outputs from each chain are indistinguishable by examining the scale reduction factor, considering values below 1.1 as an indication of convergence.

We also evaluate the quality of mixing using the ratio
$$
\mbox{Mix \%} = 100 \times \frac{\mbox{ESS}}{N}
$$
where $N$ represents the total number of samples. We use 1,000 burn-in samples followed by 5,000 samples for inference. The computations were performed on an Apple M1 Pro with 12 cores and 16 GB of RAM. The Gibbs sampler methods were implemented in the R programming language (version 4.4.2), leveraging the computational efficiency of the \CRANpkg{Rcpp} (version 1.0.13-1) \citep{Eddelbuettel1,Eddelbuettel2,Eddelbuettel3,Eddelbuettel4} and \CRANpkg{RcppArmadillo} (version 14.2.2-1) \citep{Eddelbuettel5,Eddelbuettel6} packages.

Table \ref{tab:ngtp_results_lasso} presents the mixing percentages, sampling efficiencies, and elapsed times (in seconds) for the modified Hans and PC Gibbs samplers applied to the benchmark datasets Diabetes${}^2$, Kakadu${}^2$, and Crime. It is important to note that the Gelman–Rubin diagnostic $\widehat{R}$ for each model parameter was below 1.01, indicating convergence, and the effective sample size (ESS) for $\vbeta$ corresponds to the median of the ESS values for the $p$-dimensional vector $\vbeta$. The results indicate that the modified Hans sampler is the most efficient for sampling $\sigma^2$ and $\lambda^2$ in the Diabetes$^2$ and Kakadu$^2$ datasets. It is also the most efficient for sampling $\vbeta$ in the Kakadu$^2$, and the second most efficient—after the modified PC sampler—for sampling $\vbeta$ in the Diabetes$^2$ dataset. Furthermore, the modified Hans sampler achieves the shortest computation time in the Diabetes$^2$ and Crime datasets. Note that \CRANpkg{monomvn} was extremely slow on the Kakadu$^2$ dataset. Therefore, its output is reported as NA, as it was not within a comparable range of performance with the other samplers.

\begin{table}[!ht]
    \centering
    \begin{tabular}{c|l|rr|rr|rr|r}
    \toprule
            &       &  $\vbeta$  & Eff                & $\sigma^2$   & Eff                  & $\lambda^2$   & Eff                      & Time \\
    Dataset & Method & Mix \%  & ($\times 10^2$)     & Mix \%  & ($\times 10^2$)       & Mix \%  & ($\times 10^2$)           &  (s) \\
\midrule

Diabetes$^2$ & Hans           &  26.3       &  3914.6       & 69.9       &  \textbf{10388.5}       & 25.2       & \textbf{3746.6}      & \textbf{1.2}  \\
         & PC             &  78.2       & \textbf{4939.9}       & 66.4       & 4191.9       & 21.7       & 1372.8       & 3.0  \\
         & monomvn        &  99.9       &  79.2       & 100.0       &  79.6       & 100.0       & 80.1       &  239.6 \\
         & bayeslm        &  11.7       &  1657.9       & 49.9       & 6587.3       &  4.0       &  539.5       &  1.4 \\
         & rstan           &  98.9       &   158.2       & 100.0      &   162.0       & 98.3 &  157.2       & 118.7 \\ 
         & bayesreg       &  76.4       &   2043.4       & 60.8       &   1627.3       & 17.8       &  477.0       &  7.0 \\ 
\midrule 

Kakadu$^2$  & Hans           &  18.9      &  \textbf{700.6}        & 73.1     & \textbf{2702.3}      &  4.8       &  \textbf{178.5}      &  5.1\\
         & PC             & 85.5        & 41.8        & 71.0      & 34.7     & 8.8      & 4.3        &  388.1 \\
         & monomvn        & NA        &  NA        & NA       &  NA        & NA       &  NA       &  NA \\
         & bayeslm        &  14.7        &  186.6        &  49.6      &  628.1      &  2.9       &   37.0     &  \textbf{3.1} \\
         & rstan           & 98.5        &  22.6       & 96.1      &  22.1       & 95.0      &  21.8       & 173.7 \\ 
         & bayesreg       & 84.3        &  190.3     & 60.5     &  136.6       & 4.2      &   9.5    &  17.7 \\ 

\midrule 
Crime    & Hans           &  2.9     &  229.3        & 72.6      & \textbf{5740.9}    &   4.7   & 377.5  &  \textbf{0.5} \\
         & PC             & 86.5    &  555.2       & 90.7       & 582.7     &  30.1    & 193.2  &   6.2 \\
         & monomvn        &  97.9  &    5.8         & 95.8       &    5.7       &  96.5       &  5.7  & 666.9  \\
         & bayeslm        &  2.8      &   142.7      & 81.8       &  4042.3       &   7.9     &  \textbf{391.0} &  0.8 \\
         & rstan           & 97.9       & 28.2        & 100.0      &    28.8        &  100.0      &  31.1  & 138.8  \\ 
         & bayesreg       & 85.9       &  \textbf{617.0}      & 80.6      &  578.9        &  29.7     &  213.6 &  5.5 \\      
         \bottomrule
    
    \end{tabular}
    \caption{Mixing percentages, efficiencies, and computation times (in seconds) for each dataset using the Hans and PC Gibbs samplers from the \pkg{BayesianLasso} package, as well as the R packages \CRANpkg{monomvn}, \CRANpkg{bayeslm}, \CRANpkg{rstan}, and \CRANpkg{bayesreg}.}
    \label{tab:ngtp_results_lasso}
\end{table}

% \begin{figure}[!ht]
%     \centering
%     \includegraphics[width=15cm, height=9.2cm]{diabetes.pdf}
%     %\vspace*{-0.5cm}
%     \caption{The densities plots display the estimated regression coefficients,
%     \texorpdfstring{$\sigma^2$, and $\lambda^2$}{sigma^2, and lambda^2} for the model fitted on the diabetes dataset using the Lasso penalty.}
%     \label{diabetes-beta-plot}
% \end{figure}

\section{Discussion}
\label{Sec:discussion}
This paper introduces the \pkg{BayesianLasso} R package, which provides a comprehensive and computationally efficient implementation of Bayesian Lasso regression based on a newly defined Lasso distribution. The package encapsulates recent methodological advances by offering implementations of both the modified Hans and PC Gibbs samplers, tailored to exploit the distributional structure of the Lasso prior.

Central to the package is the formal development of the Lasso distribution, which we establish as a member of the exponential family. We derive key theoretical properties—including the probability density function, cumulative distribution function, moments, and a numerically stable inverse-CDF sampling method—which underpin the samplers implemented in the package. In particular, the Lasso distribution serves as the full conditional distribution for regression coefficients in the proposed Gibbs sampling framework.

The \pkg{BayesianLasso} package is designed with usability and extensibility in mind, offering an accessible interface for researchers and practitioners to fit Bayesian Lasso models in regression settings where $n>p$. The package also provides utility functions for working directly with the Lasso distribution, including density evaluation, random sampling, and cumulative probability calculations.

Our empirical evaluations demonstrate that the modified Hans sampler achieves efficient mixing for sampling from $\sigma^2$ and $\lambda^2$ in the Diabetes$^2$ and Kakadu$^2$ datasets, and it also achieves efficient computational scalability in the Diabetes$^2$ and Crime datasets, making it a practical tool for posterior inference. By integrating this method into an easy-to-use R package, we aim to lower the barrier to adoption and encourage broader use of Lasso-based Bayesian modeling in applied statistical work.

\section{Code availability}

The R package \pkg{BayesianLasso} implementing the methods described in this paper is available on CRAN at \url{https://CRAN.R-project.org/package=BayesianLasso} and on GitHub at \url{https://github.com/garthtarr/BayesianLasso}.

\section{Competing interests}
The authors declare that they have no conflict of interest.

\section*{Acknowledgments}

The following source of funding is gratefully acknowledged: Australian Research Council Discovery Project grant (DP210100521).

\bibliography{references}

\address{John T. Ormerod\\
  School of Mathematics and Statistics, University of Sydney\\
  Sydney\\
  Australia\\}
\email{john.ormerod@sydney.edu.au}

\address{Mohammad Javad Davoudabadi\\
  School of Mathematics and Statistics, University of Sydney\\
  Sydney\\
  Australia\\}
\email{mohammad.davoudabadi@sydney.edu.au}

\address{Garth Tarr\\
  School of Mathematics and Statistics, University of Sydney\\
  Sydney\\
  Australia\\}
\email{garth.tarr@sydney.edu.au}

\address{Samuel Muller\\
   Faculty of Science and Engineering, Macquarie University\\
  Sydney\\
  Australia\\}
\email{samuel.muller@mq.edu.au}

\address{Jonathon Tidswell\\
  School of Mathematics and Statistics, University of Sydney\\
  Sydney\\
  Australia\\}
\email{jonathon.tidswell@sydney.edu.au}

\newpage

 \setcounter{page}{1}

\begin{center}
{\Large
Supplementary material to \\ ``The Lasso Distribution: Properties, Sampling Methods, and Applications in Bayesian Lasso Regression''}

\bigskip

by Mohammad Javad Davoudabadi, Jonathon Tidswell, Samuel Muller,  Garth Tarr \\ and John T. Ormerod

\bigskip 
\date{\today}
\end{center}

\appendix

\section{Derivation of the full conditional for each $\beta_j$}\label{Suppl:derivation_fullConditional_Dist}
Consider $p({\bf y}| \boldsymbol{\beta}, {\bf a})$, $p(\boldsymbol{\beta}| {\bf a})$, $p({\bf a})$, $p(\sigma^2)$, and $p(\lambda^2)$ are the likelihood function, and the priors of $\boldsymbol{\beta}$, ${\bf a}$, $\sigma^2$, and $\lambda^2$, respectively. The log of the full joint distribution is proportional to
\begin{align*}
    \log p(\boldsymbol{\beta}, \sigma^2, \lambda^2,
{\bf a}) & = \log \big(p({\bf y}| \boldsymbol{\beta,{\bf a}})p(\boldsymbol{\beta}|{\bf a})p({\bf a})p(\sigma^2)p(\lambda^2) \big)\\ &
\propto - \frac{1}{2\sigma^2} \Big[ {\bf y}^T{\bf y} - 2{\bf y}^T \mathbf{X}^T \boldsymbol{\beta} + \boldsymbol{\beta}^T \mathbf{X}^T \mathbf{X} \boldsymbol{\beta} + \lambda^2 \boldsymbol{\beta}^T{\bf A}\boldsymbol{\beta} \Big]  \\ 
& \quad\quad{}- \frac{\sum_{i=1}^p a_i}{2} - 2\sum_{i=1}^p a_i - \sum_{i=1}^p\frac{1}{2a_i} + \mbox{constant}; 
\end{align*}
where ${\bf a}= (a_1,...,a_p)$ and ${\bf A}= \text{diag}({\bf a}^{-1})$. After integrating out ${\bf a}$ from the above proportion, the marginal distribution of $\beta_j$ is as follows
\begin{align*}
    \log p(\beta_j) \propto - \frac{\parallel  \mathbf{X}_j\parallel^2_2}{2\sigma^2}\beta_j^2 + \frac{\mathbf{X}_j^T({\bf y} - \mathbf{X}_{-j}\boldsymbol{\beta}_{-j})}{\sigma^2} \beta_j - \frac{\lambda^2}{\sigma} |\beta_j|;
\end{align*}
which is a univariate Lasso distribution with parameters $a = \frac{\parallel  \mathbf{X}_j\parallel^2_2}{\sigma^2}$, $b = \frac{\mathbf{X}_j^T({\bf y} - \mathbf{X}_{-j}\boldsymbol{\beta}_{-j})}{\sigma^2}$, and $c = \frac{\lambda^2}{\sigma}$.

\section{Derivation of the distributional properties of the Lasso distribution}\label{Suppl:deriv_LassoDist_properties}
If $X \sim \mbox{Lasso}(a,b,c)$ with then it has density given by
$$
p(X,a,b,c) = Z^{-1}\exp\left( -\tfrac{1}{2}aX^2 + bX - c|X| \right)
$$
where $Z$ is the normalizing constant. Then
$$
\begin{array}{rl}
Z(a,b,c)
& = \int_{-\infty}^\infty \exp\left[ -\tfrac{1}{2}aX^2 + bX - c|X| \right] dx
\\ [2ex]
& 
= \int_0^\infty    \exp\left[ -\tfrac{1}{2}aX^2 + (b - c)X \right] dx
+ \int_{-\infty}^0 \exp\left[ -\tfrac{1}{2}aX^2 + (b + c)X \right] dx
\\ [2ex]
&  
= \int_0^\infty \exp\left[ - \frac{(X - \mu_1)^2}{2\sigma^2} + \frac{\mu_1^2}{2\sigma^2} \right] dx
+ \int_{-\infty}^0 \exp\left[ - \frac{(X - \mu_2)^2}{2\sigma^2} + \frac{\mu_2^2}{2\sigma^2} \right] dx
\\ [2ex]
&  
= \sqrt{2\pi\sigma^2}
\left[  \exp\left\{  \frac{\mu_1^2}{2\sigma^2} \right\} \int_0^\infty \phi(x;\mu_1,\sigma^2) dx
+       \exp\left\{  \frac{\mu_2^2}{2\sigma^2} \right\} \int_{-\infty}^0 \phi(x;\mu_2,\sigma^2) dx
\right] 
\\ [2ex]
&  
= \sqrt{2\pi\sigma^2}
\left[  \exp\left\{  \frac{\mu_1^2}{2\sigma^2} \right\} \left\{ 1 - \Phi(-\mu_1/\sigma) \right\} 
+       \exp\left\{  \frac{\mu_2^2}{2\sigma^2} \right\} \left\{\Phi(-\mu_2/\sigma) \right\} 
\right] 
\\ [2ex]
&  
= \sqrt{2\pi\sigma^2}
\left[  \exp\left(  \frac{\mu_1^2}{2\sigma^2} \right) \Phi\left(\frac{\mu_1}{\sigma} \right) 
+       \exp\left(  \frac{\mu_2^2}{2\sigma^2} \right) \Phi\left( \frac{-\mu_2}{\sigma} \right)  
\right] 
\\ [2ex]
& 
= 
  \sigma \left[ \frac{\Phi(\mu_1/\sigma)}{\phi(\mu_1/\sigma)}
+ \frac{\Phi(-\mu_2/\sigma)}{\phi(-\mu_2/\sigma)}  \right] 
\end{array} 
$$
where $\mu_1 = (b-c)/a$, $\mu_2 = (b + c)/a$ and $\sigma^2 = 1/a$.

The MGF of the Lasso distribution is
\begin{align*}
    M(t) = \mathbb{E} (\exp (tX)) & = Z(a,b,c)^{-1} \int \exp (tX) \exp (\frac{-a}{2}X^2 + bX -c|X|) dx \\ &
    = Z(a,b,c)^{-1} \int \exp (\frac{-a}{2}X^2 + (b+t) X -c|X|) dx \\ &
    = \frac{Z(a,b+t,c)}{Z(a,b,c)}Z(a,b+t,c)^{-1} \int \exp (\frac{-a}{2}X^2 + (b+t) X -c|X|) dx  \\ &
    = \frac{Z(a,b+t,c)}{Z(a,b,c)}.
\end{align*}

The moments of the Lasso distribution are:
$$
\begin{array}{rl}
\mathbb{E}(X^r)
& = Z^{-1} \int_{-\infty}^\infty X^r \exp\left[ -\tfrac{1}{2}aX^2 + bX - c|X| \right] dx
\\ [2ex]
&  
= Z^{-1}  \int_0^\infty   X^r \exp\left[ -\tfrac{1}{2}aX^2 + (b - c)X \right] dx
+ \int_{-\infty}^0 X^r \exp\left[ -\tfrac{1}{2}aX^2 + (b + c)X \right] dx
\\ [2ex]
&  
= Z^{-1}  \sqrt{2\pi\sigma^2}
 \exp\left(  \frac{\mu_1^2}{2\sigma^2} \right) \int_0^\infty X^r \phi(x;\mu_1,\sigma^2) dx
 +   \sqrt{2\pi\sigma^2}   \exp\left(  \frac{\mu_2^2}{2\sigma^2} \right) \int_{-\infty}^0 X^r \phi(x;\mu_2,\sigma^2) dx

\\ [2ex]
& 
= \frac{\sigma}{Z} \left[  
 \frac{\Phi(\mu_1/\sigma)}{\phi(\mu_1/\sigma)} \frac{\int_0^\infty X^r \phi(x;\mu_1,\sigma^2) dx}{\Phi(\mu_1/\sigma)}
+  \frac{\Phi(-\mu_2/\sigma)}{\phi(-\mu_2/\sigma)}  \frac{\int_{-\infty}^0 X^r \phi(x;\mu_2,\sigma^2) dx}{\Phi(\mu_2/\sigma)}
\right] 

\\ [4ex]
&  
= \frac{\sigma}{Z} \left[  
 \frac{\Phi(\mu_1/\sigma)}{\phi(\mu_1/\sigma)} 
 \mathbb{E}( A^r )
+ \frac{\Phi(-\mu_2/\sigma)}{\phi(-\mu_2/\sigma)}  \mathbb{E}( B^r )
\right] 

\\ [4ex]
&  
= \frac{\sigma}{\sigma \left[ \frac{\Phi(\mu_1/\sigma)}{\phi(\mu_1/\sigma)}
+ \frac{\Phi(-\mu_2/\sigma)}{\phi(-\mu_2/\sigma)}  \right]} \left[  
 \frac{\Phi(\mu_1/\sigma)}{\phi(\mu_1/\sigma)} 
 \mathbb{E}( A^r )
+  \frac{\Phi(-\mu_2/\sigma)}{\phi(-\mu_2/\sigma)}  \mathbb{E}( B^r )
\right] 
\\ [4ex]
&  
= \left[ \frac{\frac{\Phi(\mu_1/\sigma)}{\phi(\mu_1/\sigma)}}{\left[ \frac{\Phi(\mu_1/\sigma)}{\phi(\mu_1/\sigma)}
+ \frac{\Phi(-\mu_2/\sigma)}{\phi(-\mu_2/\sigma)}  \right]}    
 \mathbb{E}( A^r )
+  \frac{\frac{\Phi(-\mu_2/\sigma)}{\phi(-\mu_2/\sigma)}}{\left[ \frac{\Phi(\mu_1/\sigma)}{\phi(\mu_1/\sigma)}
+ \frac{\Phi(-\mu_2/\sigma)}{\phi(-\mu_2/\sigma)}  \right]}   \mathbb{E}( B^r )
\right] 
\\ [4ex]
&  
=
 (1 - w) \mathbb{E}(A^r) +  w \mathbb{E}(B^r)

\end{array} 
$$
where $A\sim TN_+(\mu_1,\sigma^2)$, $B\sim TN_-(\mu_2,\sigma^2)$ and $TN_+$ is denotes the positively truncated normal distribution.

If $x\le 0$ the CDF is given by
$$
\begin{array}{rl} 
P(X < x) 
&  = Z^{-1} \int_{-\infty}^x \exp\left[ -\tfrac{1}{2}at^2 + (b + c)t \right] dt 
\\ [2ex]
& = Z^{-1} \sqrt{2\pi\sigma^2} \exp\left( \frac{\mu_2^2}{2\sigma^2} \right) \int_{-\infty}^x \phi(t;\mu_2,\sigma^2) dt
\\ [2ex]
& = Z^{-1} \sqrt{2\pi\sigma^2} \exp\left( \frac{\mu_2^2}{2\sigma^2} \right) \Phi\left( \frac{x - \mu_2}{\sigma} \right)
\\ [2ex]
& =  \frac{\sigma}{Z}\frac{ \Phi\left( \frac{x - \mu_2}{\sigma} \right)}{\phi(-\mu_2/\sigma)}
\\ [2ex]
& = w\frac{\Phi\left( 
\frac{x - \mu_2}{\sigma}\right)}{\Phi(-\mu_2/\sigma)} 

\end{array} 
$$
and if $x>0$ we have
$$
\begin{array}{rl}
P(X < x)
&  = Z^{-1}\int_{-\infty}^x \exp\left( -\tfrac{1}{2}at^2 + bt - c|t| \right) dt
\\ [2ex]
& 
= Z^{-1} \sqrt{2\pi\sigma^2}
\left[  \exp\left(  \frac{\mu_1^2}{2\sigma^2} \right) \int_0^x \phi(t;\mu_1,\sigma^2) dt
+      \exp\left(  \frac{\mu_2^2}{2\sigma^2} \right) \Phi\left( \frac{-\mu_2}{\sigma} \right)  
\right] 
\\ [2ex]
&  = \frac{\sigma}{Z} 
\left[  \frac{ 
\Phi\left( \frac{x - \mu_1}{\sigma} \right) - \Phi\left( \frac{ - \mu_1}{\sigma} \right)}{\phi(\mu_1/\sigma)}
+       \frac{\Phi\left( -\mu_2/\sigma \right)}{\phi(-\mu_2/\sigma)}  
\right] 
\\ [2ex]
&  = \displaystyle 1 - (1 - w)\frac{\Phi(\frac{\mu_1 - x}{\sigma})}{\Phi(\mu_1/\sigma)}.

\end{array} 
$$
For the inverse CDF, we again have two cases. Let $u = P(X<x)$. When 
$$
u \le \frac{\sigma}{Z}  \frac{ \Phi\left(  - \mu_2/\sigma \right)}{\phi(-\mu_2/\sigma)} = w
$$ 
we solve
$$
u = \frac{ \sigma \Phi\left( \frac{x - \mu_2}{\sigma} \right)}{Z \phi(-\mu_2/\sigma)}
$$
for $x$ to obtain
$$
x = \mu_2 + \sigma \,\Phi^{-1}\left[ (Z/\sigma) \phi(-\mu_2/\sigma) u \right] = \mu_2 + \sigma \,\Phi^{-1}\left[ \Phi(-\mu_2/\sigma) u /w\right].
$$ 
When 
$$
u > \frac{\sigma}{Z}  \frac{ \Phi\left(  - \mu_2/\sigma \right)}{\phi(-\mu_2/\sigma)} = w
$$ 
we need to solve
$$
u = \displaystyle 1 - (1 - w)\frac{\Phi(\frac{\mu_1 - x}{\sigma})}{\Phi(\mu_1/\sigma)}
$$
for $x$ to obtain
$$
x = \mu_1 - \sigma\, \Phi^{-1}\left[ \Phi(\mu_1/\sigma) (1-u)/(1-w) \right]. 
$$

The Lasso distribution belongs to the exponential family distributions based on the following in which ${\boldsymbol\theta} = (a, b, c)$ is the set of parameters
\begin{align*}
p(X|{\boldsymbol\theta})
    & = h(X)\exp\left[{\boldsymbol\nu}({\boldsymbol\theta}) \textbf{T}(X)-\mathbb{A}({\boldsymbol\theta})\right]\\ 
%    & =Z^{-1}\exp(-\frac{1}{2}ax^2+bx-c|x|)\\&
%    =\sigma^* \left[\frac{\Phi(\mu_1/\sigma^*)}{\phi(\mu_1/\sigma^*)} + \frac{\Phi(-\mu_2/\sigma^*)}{\phi(-\mu_2/\sigma^*)} \right]^{-1} \exp(-\frac{1}{2}ax^2+bx-c|x|)\\&
    & = \exp\left[ -\tfrac{1}{2}aX^2+bX-c|X| - \left\{ \log\left(\frac{\Phi(\mu_1/\sigma)}{\phi(\mu_1/\sigma)} + \frac{\Phi(-\mu_2/\sigma)}{\phi(-\mu_2/\sigma)} \right) + \log (\sigma)  \right\} \right], 
\end{align*}
where
$$
\begin{array}{c}
{\boldsymbol\nu}({\boldsymbol\theta}) 
    = (-\frac{1}{2}a,b,-c), 
    \quad
h(X) = 1, 
    \quad 
\textbf{T}(X) = (X^2,X,|X|), \quad \text{and}
\\
\displaystyle \mathbb{A}({\boldsymbol\theta}) 
    = \log\left[ \frac{\Phi(\mu_1/\sigma)}{\phi(\mu_1/\sigma)} + \frac{\Phi(-\mu_2/\sigma)}{\phi(-\mu_2/\sigma)} \right] + \log (\sigma).
\end{array}
$$

%% file: RJwrapper.bbl
\begin{thebibliography}{29}
\providecommand{\natexlab}[1]{#1}
\providecommand{\url}[1]{\texttt{#1}}
\expandafter\ifx\csname urlstyle\endcsname\relax
  \providecommand{\doi}[1]{doi: #1}\else
  \providecommand{\doi}{doi: \begingroup \urlstyle{rm}\Url}\fi

\bibitem[Bürkner et~al.(2023)Bürkner, Gabry, Kay, and Vehtari]{BurknerEtAl2023}
P.-C. Bürkner, J.~Gabry, M.~Kay, and A.~Vehtari.
\newblock posterior: Tools for working with posterior distributions, 2023.
\newblock URL \url{https://mc-stan.org/posterior/}.
\newblock R package version 1.4.1.

\bibitem[Croissant and Graves(2022)]{Ecdat}
Y.~Croissant and S.~Graves.
\newblock \emph{Ecdat: Data Sets for Econometrics}, 2022.
\newblock URL \url{https://CRAN.R-project.org/package=Ecdat}.
\newblock R package version 0.4-2.

\bibitem[Eddelbuettel(2013)]{Eddelbuettel3}
D.~Eddelbuettel.
\newblock \emph{Seamless {R} and {C++} Integration with {Rcpp}}.
\newblock Springer, New York, 2013.
\newblock \doi{10.1007/978-1-4614-6868-4}.
\newblock ISBN 978-1-4614-6867-7.

\bibitem[Eddelbuettel and Balamuta(2018)]{Eddelbuettel4}
D.~Eddelbuettel and J.~J. Balamuta.
\newblock {Extending {R} with {C++}: A Brief Introduction to {Rcpp}}.
\newblock \emph{The American Statistician}, 72\penalty0 (1):\penalty0 28--36, 2018.
\newblock \doi{10.1080/00031305.2017.1375990}.

\bibitem[Eddelbuettel and Fran\c{c}ois(2011)]{Eddelbuettel2}
D.~Eddelbuettel and R.~Fran\c{c}ois.
\newblock {Rcpp}: Seamless {R} and {C++} integration.
\newblock \emph{Journal of Statistical Software}, 40\penalty0 (8):\penalty0 1--18, 2011.
\newblock \doi{10.18637/jss.v040.i08}.

\bibitem[Eddelbuettel and Sanderson(2014)]{Eddelbuettel5}
D.~Eddelbuettel and C.~Sanderson.
\newblock Rcpparmadillo: Accelerating {R} with high-performance {C}++ linear algebra.
\newblock \emph{Computational Statistics and Data Analysis}, 71:\penalty0 1054--1063, March 2014.
\newblock \doi{10.1016/j.csda.2013.02.005}.

\bibitem[Eddelbuettel et~al.(2024{\natexlab{a}})Eddelbuettel, Francois, Allaire, Ushey, Kou, Russell, Ucar, Bates, and Chambers]{Eddelbuettel1}
D.~Eddelbuettel, R.~Francois, J.~Allaire, K.~Ushey, Q.~Kou, N.~Russell, I.~Ucar, D.~Bates, and J.~Chambers.
\newblock \emph{Rcpp: Seamless {R} and C++ Integration}, 2024{\natexlab{a}}.
\newblock URL \url{https://CRAN.R-project.org/package=Rcpp}.
\newblock R package version 1.0.13-1.

\bibitem[Eddelbuettel et~al.(2024{\natexlab{b}})Eddelbuettel, Francois, Bates, Ni, and Sanderson]{Eddelbuettel6}
D.~Eddelbuettel, R.~Francois, D.~Bates, B.~Ni, and C.~Sanderson.
\newblock \emph{RcppArmadillo: 'Rcpp' Integration for the 'Armadillo' Templated Linear Algebra Library}, 2024{\natexlab{b}}.
\newblock URL \url{https://CRAN.R-project.org/package=RcppArmadillo}.
\newblock R package version 14.2.2-1.

\bibitem[Efron et~al.(2004)Efron, Hastie, Johnstone, and Tibshirani]{efron2004least}
B.~Efron, T.~Hastie, I.~Johnstone, and R.~Tibshirani.
\newblock Least angle regression.
\newblock \emph{The Annals of Statistics}, 32\penalty0 (2):\penalty0 407 -- 499, 2004.
\newblock \doi{10.1214/009053604000000067}.
\newblock URL \url{https://doi.org/10.1214/009053604000000067}.

\bibitem[Gelman and Rubin(1992)]{gelman1992inference}
A.~Gelman and D.~B. Rubin.
\newblock Inference from iterative simulation using multiple sequences.
\newblock \emph{Statistical science}, 7\penalty0 (4):\penalty0 457--472, 1992.
\newblock URL \url{https://doi.org/10.1214/ss/1177011136}.

\bibitem[Golub and van Loan(2013)]{golub13}
G.~H. Golub and C.~F. van Loan.
\newblock \emph{Matrix Computations}.
\newblock The Johns Hopkins University Press, fourth edition, 2013.
\newblock ISBN 9781421407944.

\bibitem[Gramacy(2024)]{monomvn}
R.~B. Gramacy.
\newblock \emph{monomvn: Estimation for MVN and Student-t Data with Monotone Missingness}, 2024.
\newblock URL \url{https://CRAN.R-project.org/package=monomvn}.
\newblock R package version 1.9-21.

\bibitem[Hans(2009)]{hans2009bayesian}
C.~Hans.
\newblock Bayesian lasso regression.
\newblock \emph{Biometrika}, 96\penalty0 (4):\penalty0 835--845, 2009.
\newblock URL \url{https://doi.org/10.1093/biomet/asp047}.

\bibitem[Hastie and Efron(2022)]{lars}
T.~Hastie and B.~Efron.
\newblock \emph{lars: Least Angle Regression, Lasso and Forward Stagewise}, 2022.
\newblock URL \url{https://CRAN.R-project.org/package=lars}.
\newblock R package version 1.3.

\bibitem[Hastie et~al.(2015)Hastie, Tibshirani, and Wainwright]{HasteEtAl2015}
T.~Hastie, R.~Tibshirani, and M.~Wainwright.
\newblock \emph{Statistical Learning with Sparsity: The Lasso and Generalizations}.
\newblock Chapman \& Hall/CRC, 2015.
\newblock ISBN 1498712169.

\bibitem[He et~al.(2022)He, Hahn, Lopes, and Herren]{bayeslm}
J.~He, P.~R. Hahn, H.~Lopes, and A.~Herren.
\newblock \emph{bayeslm: Efficient Sampling for Gaussian Linear Regression with Arbitrary Priors}, 2022.
\newblock URL \url{https://CRAN.R-project.org/package=bayeslm}.
\newblock R package version 1.0.1.

\bibitem[Makalic and Schmidt(2016)]{makalic2016high}
E.~Makalic and D.~F. Schmidt.
\newblock High-dimensional {Bayesian} regularised regression with the bayesreg package.
\newblock \emph{arXiv preprint arXiv:1611.06649}, 2016.

\bibitem[Marsaglia(2004)]{marsaglia2004evaluating}
G.~Marsaglia.
\newblock Evaluating the normal distribution.
\newblock \emph{Journal of Statistical Software}, 11:\penalty0 1--11, 2004.

\bibitem[Mills(1926)]{Mills1926}
J.~P. Mills.
\newblock Table of the ratio: Area to bounding ordinate, for any protion of normal curve.
\newblock \emph{Biometrika}, 18\penalty0 (3-4):\penalty0 395--400, 1926.

\bibitem[Mächler(2022)]{Martin}
M.~Mächler.
\newblock {Asymptotic Tail Formulas For Gaussian Quantiles}, 2022.
\newblock URL \url{https://cran.r-project.org/package=DPQ/vignettes/qnorm-asymp.pdf}.

\bibitem[Neal(2003)]{neal2003slice}
R.~M. Neal.
\newblock Slice sampling.
\newblock \emph{The Annals of Statistics}, 31\penalty0 (3):\penalty0 705--767, 2003.

\bibitem[Ormerod et~al.(2025)Ormerod, Davoudabadi, Tarr, and Mueller]{BayesianLassoPkg}
J.~Ormerod, M.~J. Davoudabadi, G.~Tarr, and S.~Mueller.
\newblock {BayesianLasso}: {Bayesian Lasso Regression and Tools for the Lasso Distribution}, 2025.
\newblock URL \url{https://github.com/garthtarr/BayesianLasso}.
\newblock R package on GitHub.

\bibitem[Park and Casella(2008)]{park2008bayesian}
T.~Park and G.~Casella.
\newblock The {B}ayesian {L}asso.
\newblock \emph{Journal of the American Statistical Association}, 103\penalty0 (482):\penalty0 681--686, 2008.

\bibitem[Redmond(2002)]{communities_and_crime_183}
M.~Redmond.
\newblock {Communities and Crime}.
\newblock UCI Machine Learning Repository, 2002.
\newblock URL \url{https://doi.org/10.24432/C53W3X}.

\bibitem[Reemtsen(1990)]{reemtsen1990modifications}
R.~Reemtsen.
\newblock Modifications of the first {R}emez algorithm.
\newblock \emph{SIAM Journal on Numerical Analysis}, 27\penalty0 (2):\penalty0 507--518, 1990.

\bibitem[{Stan Development Team}(2020)]{Stan_Development_Team2020-sz}
{Stan Development Team}.
\newblock {{RStan}}: the {R} interface to {Stan}, 2020.
\newblock URL \url{http://mc-stan.org/}.

\bibitem[{Statisticat} and {LLC.}(2021)]{LaplacesDemon}
{Statisticat} and {LLC.}
\newblock \emph{LaplacesDemon: Complete Environment for {Bayesian} Inference}, 2021.
\newblock URL \url{https://web.archive.org/web/20150206004624/http://www.bayesian-inference.com/software}.
\newblock R package version 16.1.6.

\bibitem[Tibshirani(1996)]{tibshirani1996regression}
R.~Tibshirani.
\newblock Regression shrinkage and selection via the lasso.
\newblock \emph{Journal of the Royal Statistical Society Series B: Statistical Methodology}, 58\penalty0 (1):\penalty0 267--288, 1996.

\bibitem[Vehtari et~al.(2021)Vehtari, Gelman, Simpson, Carpenter, and Bürkner]{VehtariEtAl2021}
A.~Vehtari, A.~Gelman, D.~Simpson, B.~Carpenter, and P.-C. Bürkner.
\newblock Rank-normalization, folding, and localization: An improved rhat for assessing convergence of {MCMC} (with discussion).
\newblock \emph{Bayesian Analysis}, 2021.

\end{thebibliography}
